\documentclass[pra,aps,twocolumn,epsfig,nofootinbib,showkeys,showpacs]{revtex4}
\usepackage{amsmath}
\usepackage{amssymb}
\usepackage[dvips]{graphicx}
\usepackage{dcolumn}
\usepackage{bm}
\usepackage[colorlinks=false,dvipdfm]{hyperref} 
\usepackage{setspace}
\usepackage{amsfonts}
\usepackage{rotating}
\usepackage{makeidx}
\usepackage{amsthm}

\newcommand{\tr}{\mbox{Tr}}

\begin{document}

\title{Perturbative treatment for stationary state of local master equation \footnote{Commun. Theor. Phys. 70 (2018) 38-42}}

\author{Jian-Ying Du}
\affiliation{Department of Physics, School of Science, Tianjin University, Tianjin 300072, China}

\author{Fu-Lin Zhang}
\email[Corresponding author: ]{flzhang@tju.edu.cn}
\affiliation{Department of Physics, School of Science, Tianjin University, Tianjin 300072, China}

\date{\today}

\begin{abstract}
The local approach to construct master equation for a composite open system with a weak internal coupling is simple and seems reasonable.
However, it is thermodynamic consistent only when the subsystems are resonantly coupled.
Efforts are being made to understand the inconsistency  and test the validity of the local master equation.
We present a perturbative method to solve the steady-state solutions of linear local master equations, which are  demonstrated by two simple models.
The solving process shows the stationary state as the result of competition between incoherent operations and the unitary creating quantum coherence, and consequently relate quantum coherence with  thermodynamic consistency.
\end{abstract}



\maketitle

\section{Introduction}\label{Intro}

It is generally impossible to isolate a quantum system from its surroundings,  which is referred to an open quantum
system.
Studying open quantum systems has both theoretical and practical significance, such as in quantum thermodynamics \cite{Book2004,Book2009} and quantum information \cite{Book}.
However, only few system-environment models can be solved exactly.
For most cases, the effects of environment are treated by employing various effective models, e.g., time-dependent Hamiltonian, quantum Langevin and stochastic Schr\"{ o}dinger equations, quantum state diffusion models or Hilbert-space averaging methods \cite{Berry,gisin1992quantum,gemmer2006finite,BookOpenSys,gardiner2004quantum,RMP2017dynamics}.

The most widely applied approach is the quantum master equation \cite{BookOpenSys,gardiner2004quantum,RMP2017dynamics}, which can be derived from a quantum  system-environment model by the partial trace over the environment under appropriate approximations.
The most common approximations are the Born-Markov (factorization and memoryless)  approximations and the rotating wave approximations, which  lead to the Gorini-Kossakowski-Lindblad-Sudarshan (GKLS) quantum master equation \cite{lindblad1976generators,gorini1976completely}.


The \textit{local approach} is an approximation often invoked in building the quantum master equation of a composite open system with a weak internal coupling, whose subsystems coupled to local environments.
The internal coupling is supposed to have no effect on the local dynamical generator, often in the GKLS form, of each individual component.
And the master equation is constructed by simply adding the local dynamical generators.
The other extreme is the \textit{global approach}, in which the system with a strong internal coupling is considered as a whole, and the global dynamical generators are derived in the standard procedure to build the quantum master equation.
Each of the two approaches has its own pros and cons.
The local GKLS master equation is found to be thermodynamic inconsistent, which may lead to a heat current flowing against the temperature gradient \cite{EPL2014local}.
But, in deriving the global one, the eigenvalue problem for the system Hamiltonian can be difficult.
In addition, for a global GKLS master equation, the steady-state, which is a mixture of eigenstates of the whole system,  is often far from direct product states of subsystems.
This makes it to be hard to define the local temperature of a subsystem \cite{PRX2014locality}.
 Currently, efforts are being made to study the  local and  global GKLS master equations and test their validity or divergence \cite{manrique2015nonequilibrium,EPL2016perturbative,hofer2017markovian,gonzalez2017testing,PRE2014re,PRE2013performance,OC2017two,man2017smallest}.


In this contribution, we are going to present a perturbative method to solve the stationary states of local master equations, with linear local dynamical generators.
The steady-state is, in general, the most important solution of a master equation, e. g. it can be regarded as the quantum counterpart of classical thermodynamic cycle in the sense of self-contained quantum thermal machines \cite{PRL2010small,PRL2012quantum,JPA2011smallest,PRE2012virtual,PRE2013performance,PRE2014re,PRE2014entanglement,SR2014quantum,ARPC2014quantum,PRE2015small,JPA2016,man2017smallest}.

On the one hand, the perturbative method further simplify the task to derive the stationary state, which is represented by a series of the strength of weak internal interaction.
And on the other hand,  the recurrence relation of the steady-state clearly shows the competition between the internal interaction and the  trend back to the local stationary state.
Particularly, when the local dynamical generators are in the GKLS form,  such competition becomes the one between incoherent operations and the unitary creating quantum coherence\cite{baumgratz2014quantifying,hu2017quantum}.
This provides a possible  perspective to relate the thermodynamic consistency with quantum coherence.
As two examples, we study the two-qubit model analysed in \cite{EPL2014local,manrique2015nonequilibrium,EPL2016perturbative} and the three-qubit absorption refrigerator \cite{PRL2010small,JPA2011smallest} to demonstrate our  perturbative method  and discuss the relation between quantum coherence and  thermodynamic consistency.


\section{Perturbative method }

In the local approach, the master equation for an open system reads
\begin{equation}\label{MasterEq}
\dot{\rho}=-i[H,\rho]+ \sum_{i} D_i{(\rho)},
\end{equation}
where $H$ is the total Hamiltonian of the whole system, and $D_i$ is the local dynamical generator on $i$th subsystem induced by the coupling with its environment.
The Hamiltonian can be written as the sum of free Hamiltonian of subsystems and internal interaction as
\begin{equation}
H=\sum_i H_i + g X,
\end{equation}
where we denote the interaction $H_{\rm int}=g X$ with $g$ being a small dimensionless constant and $X$ is a nonlocal operator specifying the couplings.

We assume that the steady-state solution of master equation (\ref{MasterEq}) exists and is unique, satisfying
\begin{equation}\label{Steadystate}
     0=-i[H,\rho_s]+ \sum_{i} D_i{(\rho_s)}.
\end{equation}
When internal interaction $g=0$, the solution is simply the direct product of local steady states
\begin{equation}\label{zeroorder}
\rho_s=\rho^{(0)}=\otimes_i \tau_i.
\end{equation}
Here the local state $\tau_i$ of a subsystem is determined by
\begin{equation}\label{local}
0=-i[H_i,\tau_i]+  D_i{(\tau_i)}.
\end{equation}




Similar with the perturbation theory described in every textbook on quantum mechanics, we represent the steady state by the series
\begin{eqnarray}\label{rhos}
\rho_{s} =  \rho^{(0)}+g \rho^{(1)}+g^2 \rho^{(2)}+...
\end{eqnarray}
 The zeroth-order term $\rho^{(0)}$ is given by Eq. (\ref{zeroorder}), and the normalization condition requires $\tr  \rho^{(k)}=0 $ for $k>0$.
When the local dynamical generators $D_i$ are linear, one can insert the series into Eq. (\ref{Steadystate}) and obtain the recurrence relation
\begin{equation}\label{EQ.approximation}
     i [X,\rho^{(k)}]=-i [H^{(0)},\rho^{(k+1)}]+ \sum_{i} D_i{(\rho^{(k+1)})},
\end{equation}
where $H^{(0)}=\sum_i H_i$ is the free Hamiltonian and $k=0,1,2...$.
The steady-state problem described by Eq. (\ref{local})  is often trivial, e. g. $\tau_i$ being a thermal state in the temperature of its bath in an equilibrium
state.
Then, the task becomes to derive $\rho^{(k+1)}$ from the commutator $[X,\rho^{(k)}]$ and the properties of $H^{(0)}$ and $D_i$, starting with $\rho^{(0)}$.

When $D_i$ are in the GKLS form, the local steady states $\tau_i$ are functions of free Hamiltonian $H_i$ and thus are incoherent states with respect to the  representation of $H_i$.
And, both the two terms in the right hand of Eq. (\ref{EQ.approximation}) corresponds to the changes of $\rho^{(k+1)}$ under incoherent operations in an infinitesimal interval of time, in which the commutator originates from the unitary generated by $H_0$ and $D_i$ from the transition and dephasing caused by baths.
Whereas, the unitary deriving  by the interaction $H_{\rm int}$ may produce coherence, when it does not commute with the diagnalized states it acting on.
In such case, the recurrence relation (\ref{EQ.approximation}) shows the fact that,  the coherence generated from $\rho^{(k)}$  by the internal interaction is counteracted by the decoherence of $\rho^{(k+1)}$.
Hence, the steady-state solution is  the result of competition between the internal interaction and incoherent operations.

One can simply rewrap the density matrix $\rho_s$ to a vector, and simultaneously $D_i$ and $-i[H,\bullet]$ to matrices operating on it, as the treatments of stationary state in \cite{PRE2014re} and transient state in \cite{PRE2015small}.
In this way, the series $\rho^{(k)}$ can be derived  by using the perturbation theory for algebraic eigenvalue problem \cite{wilkinson1965algebraic}.
In present work, we omit this standard method, but show a process for construction of the series of steady state by using two simple examples in the following parts.
In these examples, the generating and destroying of coherence in Eq. (\ref{EQ.approximation}) are shown visually.


 \section{Two-qubit heat transfer network}

The first example is the simplest heat transfer network model composed of two qubits, $1$ and $2$,  each of which is coupled to a single heat bath with temperature $T_1$ and $T_2$ \cite{EPL2014local,manrique2015nonequilibrium,EPL2016perturbative}.
Let us denote the Pauli operators for qubit $i$ as $\sigma_i^{x,y,z}$ and $\sigma_i^{\pm}=( \sigma_i^{x } \pm i \sigma_i^{y} )/2$.
The free Hamiltonian is given by
\begin{eqnarray}\label{Hi}
H_i=E_{i} \frac{\sigma_i^{z}}{2}.
\end{eqnarray}
And the two subsystems are weakly coupled to each other with the bipartite operator
\begin{eqnarray}
X= \sigma_1^{+} \sigma_2^{-}+\sigma_1^{-} \sigma_2^{+}.
\end{eqnarray}
The local steady state $\tau_i$ is a thermal state,
\begin{eqnarray}\label{taui}
\tau_{i}=\frac{1}{2}(1+s_{i}\sigma_i^{z}),
\end{eqnarray}
where $s_{i}=\tanh(-\beta_i E_i/2)$ with $\beta_i=1/T_i$.
For simplify, we model the local dissipator for each bath on its corresponding qubit as \cite{PRL2010small}
\begin{eqnarray}\label{Di}
 D_i{(\rho)}=p_{i}\bigr (\tau_{i}\otimes \tr_{i}\rho-\rho\bigr),
 \end{eqnarray}
where $p_i$ is the dissipation rate, depending on how well each qubit is relative to its bath.
It is a modified version of the one derived from the Jaynes-Cummings (JC) model with the dephasing rate being doubled \cite{du2017nonequilibrium}, and can be rewritten explicitly in the GKLS form (pointed out in the supplementary material of \cite{PRL2010small}).





Substituting the zero-order steady state $\rho^{(0)} =\tau_{1}\otimes \tau_{2}$ and the interaction operator into Eq. (\ref{EQ.approximation}), one can obtain
\begin{equation}
i[X,\rho^{(0)}]=   \Delta s Y,
\end{equation}
where $\Delta s= (s_1-s_2)/2$ and
\begin{equation}
Y=-i \sigma_1^+ \sigma_2^- + i \sigma_1^- \sigma_2^+.
\end{equation}
It is obvious that the infinitesimal interaction generates the coherence term $Y$.
One can directly  assume the first order $\rho^{(1)}$ consists of $Y$, which is suppressed by $D_i$.
In addition,  the commutators $[H^{(0)}, Y] \propto \Delta E X$ and $[H^{(0)}, X] \propto \Delta E Y$, where $\Delta E= E_1- E_2$.
That is, the free Hamiltonian rotates the off-diagonal term in the space of  $\{X,Y\}$ when $\Delta E \neq 0$.
Therefore, one can assume
\begin{eqnarray}
\rho^{(1)} = m^{(1)}X+d^{(1)}Y,
\end{eqnarray}
where $m^{(1)}$ and $d^{(1)}$ are the parameters to be determined.
Inserting the form of $\rho^{(1)} $ into the right hand of  (\ref{EQ.approximation}), we obtain
\begin{eqnarray}
  d^{(1)} =\frac{- q \Delta s }{q^2+  \Delta E^2 },\ \ \ \
  m^{(1)} =- \frac{\Delta E }{q} d^{(1)},
\end{eqnarray}
where $q=\sum_i p_i $.

 Substituting the first-order term $\rho^{(1)}$ into the left hand of  (\ref{EQ.approximation}), the commutators $[X, \rho^{(1)}] \propto \sigma_1^z -\sigma_2^z$.
 It commutes with the free Hamiltonian $H_0$.
 According with the effects of local dissipators on $\sigma_i^z$,  we suppose
\begin{eqnarray}
\rho^{(2)} = a_1^{(2)}\sigma_1^z+a_2^{(2)}\sigma_2^z +b^{(2)}\sigma_1^z \sigma_2^z,
\end{eqnarray}
with $a_i^{(2)}$ and $b ^{(2)}$ to derive.
It is easily to obtain
\begin{eqnarray}
&& a_1^{(2)}=\frac{1}{ p_1} d^{(1)} ,\ \ \  a_2^{(2)}=-\frac{1}{ p_2} d^{(1)},  \\
 &&b^{(2)}=\frac{1}{q} \biggr( \frac{p_2 s_2}{p_1}- \frac{p_1 s_1}{p_2}\biggr)  d^{(1)},\nonumber
\end{eqnarray}
by using the relation in Eq. (\ref{EQ.approximation}).
The left hand of  Eq. (\ref{EQ.approximation}) for $k=2$ is given by
 \begin{equation}
i[X,\rho^{(2)}]=   x \Delta s   Y ,
\end{equation}
where  $x=   -2 q^2/ [ (q^2 +\Delta E^2 )p_1 p_2 ] $.
 The linearity of the right hand leads to
  \begin{eqnarray}
\rho^{(3)} = m^{(3)}X+d^{(3)}Y,
\end{eqnarray}
 and $m^{(3)}=x m^{(1)}$, $d^{(3)}= x d^{(1)}$, or in other words $\rho^{(3)} =x \rho^{(1)}$.

From the two steps to derive  $\rho^{(2)}$ and $\rho^{(3)}$ and the linearity of the recurrence relation, one can find that the series of $\rho^{(k)}$ for $k>0$ is composed of two geometric series.
That is, the steady state is given by
\begin{eqnarray}\label{rhos2}
\rho_s&=&\tau_1\otimes \tau_2 + \sum_{j=0}^{+ \infty  }  g^{2 j +1 }x^j \rho^{(1)}+ \sum_{j=0}^{+\infty  }  g^{2 j +2 }x^j \rho^{(2)}  \nonumber \\
&=& \tau_1\otimes \tau_2 +  a_1 \sigma_1^z+a_2 \sigma_2^z + b \sigma_1^z \sigma_2^z+ m X+d Y, \ \ \ \
\end{eqnarray}
where the parameters can be directly obtained by using the sum formulae of  geometric series as
\begin{eqnarray}
&& d =  \frac{g}{1-g^2x} d^{(1)}, \ \ \  m =- \frac{\Delta E }{q} d, \nonumber\\
&& a_1=\frac{g}{ p_1} d ,\ \ \  a_2=-\frac{g}{ p_2} d,   \\
&&b=\frac{1}{q} \biggr( {p_2 s_2}{a_1}+ {p_1 s_1}{a_2}\biggr).\nonumber
\end{eqnarray}


%
%
%
%
\section{Three-qubit absorption refrigerator}

We take the three-qubit model of quantum absorption refrigerator as the second example.
It is proposed in the study of the fundamental limitation on the size of thermal machines \cite{PRL2010small}, and has raised a subsequent stream of works about self-contained quantum thermal machines \cite{PRL2012quantum,JPA2011smallest,PRE2012virtual,PRE2013performance,PRE2014re,PRE2014entanglement,SR2014quantum,ARPC2014quantum,PRE2015small,JPA2016,man2017smallest}.
Similar with the two-qubit model studied above, the three qubits $1$, $2$, and $3$, interact with three baths, $1$, $2$, and $3$, at temperatures $T_1 < T_2 < T_3$ in order.
Their free Hamiltonian, local steady states and local dissipators are discribed by Eqs. (\ref{Hi}), (\ref{taui}) and (\ref{Di}) respectively.
To cool the target, qubit $1$, a interaction is introduced as
\begin{eqnarray}
H_{\rm int}^{\rm r}=g X_{\rm r} = g(\sigma_1^+ \sigma_2^- \sigma_3^+ +\sigma_1^- \sigma_2^+ \sigma_3^- ),
\end{eqnarray}
which extracts heat from the target, and dissipates it into bath $2$ through the spiral, qubit $2$.
The qubit $3$ plays the role of the engine, which gains free energy from the hot bath $3$, to drive the heat current from the target to spiral.

The commutator of the tripartite interation and the zero-order term of steady state $\rho_{\rm r}^{(0)}=\tau_{1}\otimes \tau_{2}\otimes \tau_{3}$ is proportional to the tripartite coherent term $Y_{\rm r}= -i \sigma_1^+ \sigma_2^- \sigma_3^+ +i \sigma_1^- \sigma_2^+ \sigma_3^- $.
Such term is suppressed by $D_i$ and rotated by $H_0=H_1+H_2+H_3$ to in the space of  $\{X_{\rm r},Y_{\rm r}\}$.
We derived the first order of  steady state as
 \begin{eqnarray}
\rho_{\rm r}^{(1)} = m_{\rm r}^{(1)}X_{\rm r}+d_{\rm r}^{(1)}Y_{\rm r},
\end{eqnarray}
where
\begin{eqnarray}
  d_{\rm r}^{(1)} =\frac{- q_{\rm r} \Delta s_{\rm r} }{q_{\rm r}^2+  \Delta E_{\rm r}^2 },\ \ \ \
  m_{\rm r}^{(1)} =- \frac{\Delta E_{\rm r} }{q_{\rm r}} d_{\rm r}^{(1)}.
\end{eqnarray}
Here $\Delta E_{\rm r} =E_{1}+ E_{3}-E_{2}$, $q_{\rm r} = q_1+q_2+q_3$ and $\Delta s_{\rm r} =(s_1- s_2+s_3 -s_1 s_2 s_3)/4$.
Denoting $r_i=(1+s_i)/2$ and $\bar{r}_i=(1-s_i)/2$, the parameter $\Delta s_{\rm r} =r_1 \bar{r}_2 r_3 -\bar{r}_1 r_2 \bar{r}_3$.
Using the similar steps as the case of two-qubit model, the steady state of the three-qubit absorption refrigerator can be represented by the sum of two geometric series.
By using the sum formulae of infinite geometric series, one obtains
\begin{eqnarray}\label{rho123}
 \rho_{\rm r}^{s} &=& \tau_{1}\otimes \tau_{2}\otimes \tau_{3} +d_{\rm r} Y_{\rm r}+m_{\rm r} X_{\rm r}   \nonumber \\
 &&+ \sum_i a_{i}^{\rm r}  \sigma_i^z +\sum_{ij} b_{ij}^{\rm r} \sigma_i^z\sigma_j ^z +c^{\rm r}  \sigma_1^z\sigma_2^z\sigma_3^z,  \ \ \
\end{eqnarray}
where the parameters $a_{i}^{\rm r}$, $ b_{ij}^{\rm r} $, $c^{\rm r}$ and $m_{\rm r}$ are proportional to $d_{\rm r}$, which is given by
\begin{equation}
 d_{\rm r}=\frac{-g q_{\rm r}  \triangle s_{\rm r}}
   {q_{\rm r}^{2}+\Delta E^2_{\rm r}+4 g^{2}+2 g^{2}( \sum_{i} q_{i}+\sum_{jk} q_{jk}\omega_{jk})},
\end{equation}
with $q_{i}=\frac{p_{i}}{q_{\rm r}-p_{i}}$, $q_{jk}=\frac{p_{j}q_{k}+p_{k}q_{j}}{q_{\rm r}-p_{j}-p_{k}}$ and  $\omega_{jk}=r^{\prime}_j r^{\prime}_{k} + \bar{r}^{\prime}_{j} \bar{r}^{\prime}_k$.
Here, we set $r^{\prime}_{j=1,3}=r_j$, $\bar{r}^{\prime}_{j=1,3}=\bar{r}_j$, $r^{\prime}_{2}=\bar{r}_2$ and $\bar{r}^{\prime}_{2}=r_2$.
The proportion relations are
\begin{eqnarray}
&&m_r= -\frac{d_{\rm r}}{q_{\rm r}} \Delta E_{\rm r}, \ \ \ \  a_i=(-1)^{i+1}\frac{g }{2 p_i} d_{\rm r}, \nonumber\\
&& b_{ij} = \frac{1}{q_{\rm r}} (p_i s_i a_j + p_j s_j a_i), \\
&& c=\frac{1}{q_{\rm r}}(p_1 s_1 b_{23} + p_2 s_2 b_{31}+p_3 s_3 b_{12}-\frac{1}{2} g d_{\rm r} ). \nonumber
\end{eqnarray}
When $\Delta E_r=0$, the state $ \rho_{\rm r}^{s}$  is consistent with the result in \cite{JPA2011smallest}.

\section{Thermodynamic consistency}

The steady-state solutions enable us to derive the heat currents and verify the consistency of the local master equation with thermodynamics.
The first law of thermodynamics is a conservation law of energy, which can be expressed by the steady-state heat currents $Q_i$ as \cite{JPA1979quantum}
\begin{equation}\label{firstlaw}
\sum_i Q_i =0.
\end{equation}
The second law for an isolated system is given by \cite{Entr2013quantum}
\begin{equation}\label{secondlaw}
\frac{d \mathcal{S}}{d t} = - \sum_i \frac{Q_i}{T_i} \geq 0,
\end{equation}
stating that the rate of entropy production is nonnegative.

The heat current $Q_i$ provided by bath $i$ is defined as
\begin{eqnarray}\label{Qieqution}
 Q_{i}=\tr[H {D}_i(\rho)],
 \end{eqnarray}
which is the change of energy of an open system in state $\rho$ under the influence of ${D}_i$.
It is easy to prove that the sum of all heat currents at steady state is zero, by calculating average energy of the right hand of Eq. (\ref{Steadystate}).
That is, the first law of thermodynamics in Eq. (\ref{firstlaw}) is fulfilled.
However, the second law may be violated in the cases of nonresonant, i. e. $\Delta E\neq0$ for the two-qubit model and $\Delta E_r\neq0$ for the refrigerator.

The currents of the two-qubit steady state (\ref{rhos2}) are
\begin{eqnarray}\label{Q12eqution}
 Q_i = 2 g  d\biggr[(-1)^iE_i + \frac{p_i}{q} \Delta E\biggr].
\end{eqnarray}
 When the parameter $p_i=\gamma_i [1+\exp(-\beta_i E_i)]$, this is formally consistent with the results in \cite{EPL2014local},
 and  $p_i=J_i [1+\exp(\beta_i E_i)]$ leads to the ones in  \cite{manrique2015nonequilibrium}.
 The minor difference between our results and the two mentioned references stems from  different dephasing rate.
 Substituting the currents (\ref{Q12eqution}) into the rate of entropy production in (\ref{secondlaw}), one obtains
\begin{eqnarray}\label{S12}
\frac{d \mathcal{S}}{d t} \! =\!   \xi \biggr[\!    e^{  \sum_i (\!   -\!   1)^i \beta_i E_i}   \! \!   - \! \!   1  \!  \biggr] \!
          \biggr[\!    \sum_i (\! -\! 1)^i \beta_i E_i  \! \!  + \! \!   \Delta E \sum_i \frac{p_i }{q}\beta_i   \!   \biggr] \!  ,
\end{eqnarray}
where $\xi$ is a function of all the parameters of the two-qubit model, which is always positive.
When $\Delta E=0$, it is easy to find that, the two factors in  the two square brackets of  Eq. (\ref{S12})    are of the same sign, and consequently $\frac{d \mathcal{S}}{d t} \geq 0$.
However, when $\Delta E \neq 0$, the two factors may have opposite signs, and thus $\frac{d \mathcal{S}}{d t} < 0$.
For instance, in the case of $\Delta E < 0$ and ${  \sum_i (\!   -\!   1)^i \beta_i E_i}>0 $, the sum of the two may be less than zero.
These analyses also apply to the three-qubit refrigerator in steady state (\ref{rho123}), since its currents and entropy production are also in the forms as (\ref{Q12eqution}) and (\ref{S12}), with the replacing $\Delta E\to \Delta E_{\rm r}$, $q\to q_{\rm r}$,  $d\to d_{\rm r}$ and $\xi \to \xi_{\rm r}>0$.
These results demonstrate that, the local approach is valid only under the resonance between subsystems.

Such inconsistency can be understood by comparing the heat currents (\ref{Q12eqution}) with the ones drawn by the internal interaction from subsystems.
The later are defined as the influences of the interaction on local energies, $Q_i^g =  i \tr\{  H_i [H_{\rm int}, \rho] \}$.
For the two-qubit model in steady state (\ref{rhos2}), it is easy to obtained that
\begin{eqnarray}\label{Q12g}
 Q_i^g = 2 g  d  (-1)^iE_i.
\end{eqnarray}
The results for the refrigerator have the same form, with the mentioned replacing $d\to d_{\rm r}$.
It is directly to check that, these currents fulfill the second law in (\ref{secondlaw}) but break down the first law in (\ref{firstlaw}).
Their differences with the heat currents provided by the baths (\ref{Q12eqution}) are proportional to the amounts of detuning, and  is caused by the energy allocated to global coherent terms of  the steady state.
Only when the differences vanish, the first and second laws of thermodynamics are fulfilled simultaneously.
These analyses do not rely on the compact forms of steady states in (\ref{rhos2}) and (\ref{rho123}), as the conflict can be found by using only the first and second orders of the steady states.
These results indicate an implicit assumption of the local master equation that, the global terms of a steady state are without influence upon the currents.
That is, the local approach requires that no global coherence contributing to total energy is  produced  in the competition between the internal interaction and couplings with baths.
This is similar  to the well known fact that, the laws of thermodynamics  are broken down when open systems are correlated with their environments \cite{PRE2010entanglement,NC2018}.
Furthermore, we argue that, for a subsystem in the local approach, the rest of a composite open system plays the  role of environments.
And, the requirement of resonance is similar with the fact that, only the resonant frequencies of reservoirs are involved in the standard GKLS master equation.

\section{Summary}\label{Summ}

We present a perturbative method to solve the stationary states of linear local master equations, with the internal interaction being weak enough.
This method is demonstrated by the two-qubit heat transfer network and three-qubit absorption refrigerator, in which each qubit and its bath is modeled by a simple reset model  as the treatment in \cite{PRL2010small}.
The recurrence relation shows that the stationary state is the result of competition between incoherent operations and the unitary creating quantum coherence.
Our two examples indicate that, it is required that no global coherence contributing to total energy is produced in the competition, by the thermodynamic consistency of local master equations.

In our investigation of the consistent of local master equations with thermodynamics, we did not compare the results with the open systems under other treatments as in the recent works \cite{manrique2015nonequilibrium,EPL2016perturbative,hofer2017markovian,gonzalez2017testing,PRE2014re,PRE2013performance,OC2017two,man2017smallest},
but analyze consistency of the theory by studying the heat currents drawn by the interaction and the ones provided by the baths.
Here, we argue that, the treatment in \cite{EPL2016perturbative} is not reasonable to consider a local master equation as the limit of the global one, as they are two different extremes of the internal interactions.
It would be interesting to extend  our perturbative method to the case with strong system-environment couplings, where the higher-order and  non-Markovian effects \cite{PRA2011ZhaoXY} must be taken into account.


%
%
%

\begin{acknowledgments}
This work is supported by the NSF of China (Grant No. 11675119, No. 11575125 and No. 11105097).
\end{acknowledgments}

\bibliography{PerturbLocalMEq}

\end{document}